# Ramp reversal memory in bulk crystals of 1T-TaS$_2$

Avital Fried[†,‡], Ouriel Gotesdyner [†], Irena Feldman[#], Amit Kanigel[#] and Amos Sharoni[†,‡,*]

[†] Department of Physics, Bar-Ilan University, 5290002 Ramat-Gan, Israel

[‡] Bar-Ilan Institute of Nanotechnology and Advanced Materials (BINA), 5290002 Ramat-Gan, Israel

[#] Department of Physics, Technion-Israel Institute of Technology, Haifa, 32000, Israel

*Amos.Sharoni@biu.ac.il

## ABSTRACT

The ramp reversal memory (RRM) is a non-volatile memory effect previously observed in correlated oxides exhibiting temperature-driven metal-insulator transitions (MITs). In essence, when a system displaying RRM is heated to a specific temperature within the MIT regime - where metallic and insulating domains coexist - and then cooled by reversing the temperature ramp, the resistance increases in the subsequent heating cycle. Crucially, this increase occurs only in the vicinity of the reversal temperature, indicating that the system "remembers" this temperature. However, this memory is erased in the next heating loop. While such an effect could potentially manifest in various systems, to date, it has only been reported in thin films of correlated transition metal oxides, including VO$_2$, V$_2$O$_3$, and NdNiO$_3$. In this work, we report the observation of RRM in macroscopic crystals of the layered material 1T-TaS$_2$, which undergoes an MIT near 190 K along charge-density wave transitions. Our findings provide compelling evidence that RRM is a general phenomenon, extending beyond the previously studied oxides. Interestingly, the RRM in TaS$_2$ displays significantly different characteristics: it is observed when reversing from cooling to heating (as opposed to heating to cooling), and its magnitude - representing the "strength" of the memory - is nearly an order of magnitude larger than in correlated oxides. While we discuss potential mechanisms for the RRM in TaS$_2$, a comprehensive first-principles model is still lacking. We hope that this study will prompt further investigation into the underlying mechanisms of ramp reversal memory, enhancing our understanding of this intriguing phenomenon.

The ramp reversal memory (RRM) is a recently discovered non-volatile memory effect, observed in a number of thin film transition metal oxides (TMOs) with temperature driven metal to insulator transitions (MIT).[1] The system can be made to remember one or more temperatures using a simple heating and cooling protocol. There is a well-established heuristic model for the effect, but to date there is still no accepted first-principal explanation. A major question which we address is – how general is the RRM? I.e., does it appear in systems other than TMOs? Herin, we demonstrate an RRM effect in a different material system, a bulk crystal of the transition metal dichalcogenide 1T-TaS$_2$. [2]

The RRM has been reported for three thin films of correlated TMOs: VO$_2$, V$_2$O$_3$ and NdNiO$_3$,[1,3–5] all of which exhibit a first order temperature driven MIT, with a hysteresis between the heating and cooling curves. Their temperature driven MIT is coupled to a structural transition (or deformation, in the case of NdNiO$_3$).[6–11] Other properties these systems have that are presumed necessary for the appearance of the memory effect are: (i) The systems exhibit a spatially-phase-separated state during the phase transition, where both insulating and metallic domains co-exist.[12–14] (ii) The transition temperature can be affected by various local properties, such as strain, doping and oxygen vacancies. [8,11,15–21]

The RRM was observed as a resistance change in the resistance vs temperature (R-T) curve (during heating) following a simple heating and cooling protocol. A detailed exemplification of the effect is provided in the supplementary material (SM) section S1[SM] and further detailed in previous publications. [1,3–5] In a nutshell, beginning from the low-temperature insulating phase, the system is heated to some temperature in the spatially phase-separated state, referred to as the reversal temperature – T$_R$. Then the system is cooled again, producing a so-called reversal loop (RL). In the following heating curve, which is the memory reading measurement, the R-T exhibits a higher resistance than in the non-perturbed measurement, and the change is the largest near T$_R$. The change is attributed to local increases in the transition temperature of the MIT, as will be elaborated below. Thus, to analyze the change, one can look at the temperature dependent shift in the transition temperature, ΔT, which in essence is the temperature difference between two points on the R-T curves of the perturbed and unperturbed measurement with the same resistance value, see SM section S1 and Ref [4,5] for details. It is also possible to analyze the temperature dependent normalized change in resistance, ΔR/R. When plotting ΔR/R or ΔT as function of temperature, a peak appears corresponding to the reversal temperature (T$_R$) – this is the manifestation of the RRM.

Since heating to the fully metallic state results in erasing of the memory, the following R-T heating measurement returns to the unperturbed R-T curve.

A heuristic model can capture all properties of the RRM. It postulates that if the temperature ramp reversal (from heating to cooling) is performed while the system is in the spatially phase-separated state, then the local transition temperature of the phase boundaries (between the insulating and metallic coexisting domains) increases. We refer to the phase boundaries with increased MIT transition temperature as scars. Thus, during the following resistance measurement there is a delay in the advancement of the transition, which appears as an averaged temperature delay in the transition. Hence, the measurement of ΔT is a meaningful estimation to the magnitude of the RRM effect.[4,5] Finally, when the temperature is increased above $T_R$, the previous scars are "healed", and the memory is erased. The underlying mechanism responsible for this $T_C$ shift is still not substantiated and it may alter between different systems. Mechanisms including local strain [5] and local oxygen vacancy motion [3] were suggested. Based on this model, the RRM should potentially appear in other systems with similar characteristics.

Additional properties of the RRM which are important to mention in the context of this research are: (1) The RRM appears only on the heating curve and not on the cooling curve. (2) The magnitude of the ΔT measurement increases when more reversal loops are performed (prior to measurement). (3) The effect is nonvolatile, and if the system is not heated above the reversal temperature the memory is retained. (4) It is possible to write a few memories simultaneously, as long as a previous memory is not erased by heating the sample above its reversal temperature.

In this paper, we demonstrate that the RRM effect is also present in the correlated layered material 1T-TaS$_2$. 1T-TaS$_2$ is a Van der Waals transition metal dichalcogenide, with each layer composed of a plane of hexagonally arranged Ta atoms sandwiched by two S layers and the central Ta atoms are arranged in an octahedral structure. It undergoes several temperature-driven first order charge density wave (CDW) phase transitions, which include MIT transitions of different magnitude.[22,23] In Figure 1a we plot an R-T of the crystal we measured (see SM S2 for details on synthesis[SM]). Upon cooling, 1T-TaS$_2$ transitions from an incommensurate CDW (I-phase) below 550K to a nearly commensurate CDW (N-phase) at 350K, and finally to a commensurate CDW (C-phase) below 180K, which is accompanied by a metal insulator transition with a resistance change of nearly one order of magnitude. The heating process reveals a slightly different behavior. Upon heating from the C-phase, an additional CDW phase emerges – the stripe-like nematic Triclinic CDW (T-phase), between the C-phase and N-phase. Upon further heating, the N-phase is recovered around 280K. In the C-phase, thirteen Ta atoms form a star-of-David cluster, creating a triangular superlattice.[24,25] The insulating behavior of bulk 1T-TaS$_2$ in the C-phase was attributed to a Mott localization process, however there is still debate concerning the mechanism.[26–28,29,30] The configuration of these star-of-David clusters varies for each CDW phase.[23,31] The transition between the different phases was shown to happen through phase-coexistence,[32] thus they hold an important ingredient for observation of the RRM.

The electrical properties of 1T-TaS$_2$ can be manipulated by several factors. An optical laser pulse [33,34] or an electrical pulse [35,36] applied at the low temperature insulating phase can drive the system to a metastable hidden state, resulting in a significant decrease in resistivity. This hidden state has a temperature-tunable lifetime, which can be extremely long at low temperatures. [33,34] Additionally, the interlayer stacking order can influence the electronic properties of the CDW phases. [37–39] These properties indicate there are mechanisms that can alter in a non-volatile manner the electronic properties of 1T-TaS$_2$, i.e., it holds the second ingredient needed for observing the RRM effect.

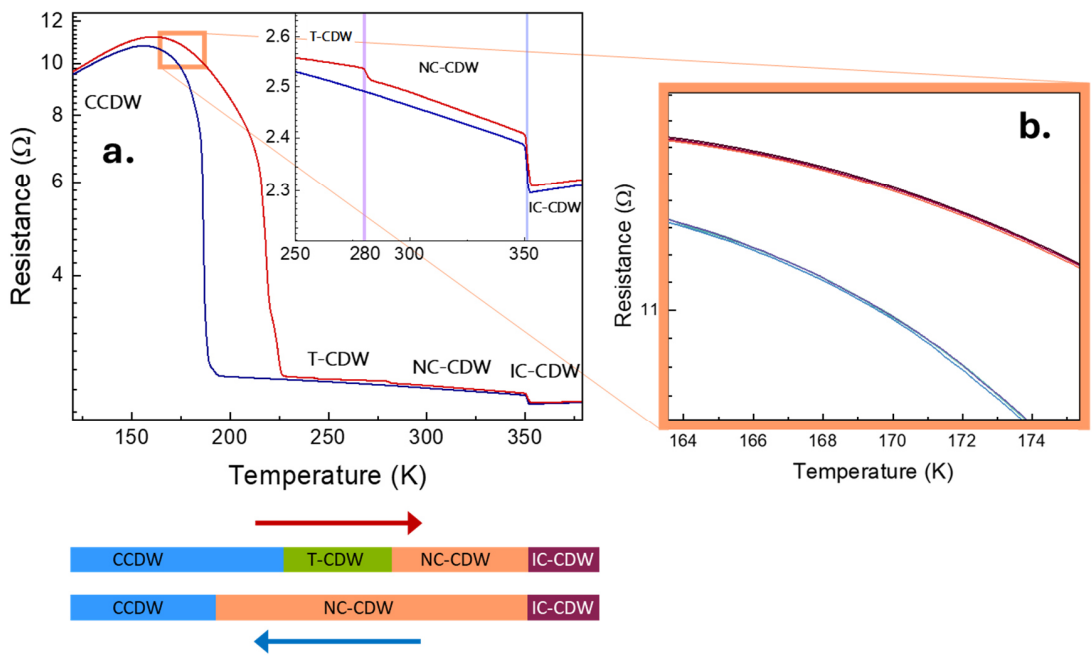

Figure 1. Properties of the 1T-TaS$_2$ crystal. (a) Resistance as a function of temperature for a bulk 1T-TaS$_2$ sample. The heating curve is shown in red and the cooling curve in blue, with the different CDW phases indicated. The inset provides a zoom-in view of the T-, N- and I- CDW phases. Below the graph: A schematic of the warm-up and cool-down CDW phase transitions of 1T-TaS2, with each phase color-coded. The start and end points of each phase align with the temperature axis in the graph. (b) R-T for four heating and cooling measurements. The heating curves are shown in different shades of red, and the cooling curves in different shades of blue. All measurements show very close values, demonstrating the reproducibility of the R-T measurements.

In this study we show there is indeed an RRM effect in bulk crystals of 1T-TaS$_2$. The main properties of the RRM are presented and analyzed, and some major differences relative to the RRM in TMOs are observed. Specifically, the RRM appears only in the cooling branch, the magnitude of the effect is nearly 10 times larger, and the memory is erased if the sample is either cooled to the C-phase or heated above the T-phase. We attempted to measure the RRM in thick flakes of 1T-TaS$_2$, but due to the metal-to-insulator transition becoming step-like it wasn't possible to see the effect.

## Results

Macroscopic millimeter-sized high-quality single crystals of 1T-TaS$_2$ were grown using the chemical vapor transport method. [40,41] Figure 1a shows the R-T measurements of one such crystal and Figure 1b shows the R-T are well repeated when performing additional R-T loops. The R-T measurements were performed in our home-made cryostat, see SM S3[54] for an image of the device and the experimental section for measurement details.

Figure 2 demonstrates the memory writing protocol of the RRM effect for a 1T-TaS$_2$ bulk sample. The RRM writing and measurement protocol for reversal temperatures on the heating curve (like those used in the study of RRM in correlated oxides [1,4]) are plotted in Figure 2a-c, and the measurements of the RRM on the cooling curve are presented in Figure 2d-g. We plot the R-T curves in Figure 2a alongside the corresponding temperature vs time plot in Figure 2b (time on the y-axis). The 'heating' measurement starts at low temperatures (i.e., the full C-phase, 140K), ramps up to the full T-phase at 240K, and then returns to the initial low temperature, forming a major loop (ML). This establishes a baseline curve for comparison with subsequent measurements (green and orange curves). Next, the sample is heated to a specific "reversal temperature" (T$_R$) of 219K, i.e., an intermediate temperature in the phase-coexistence stage (marked by a vertical gray dashed line in Figures 2a-c), and then cooled back to 140K, creating a reversal loop (RL). We performed 5 consecutive RLs, ending at 140K. Finally, another ML is performed by heating the sample to 240K and cooling back to 140K, constituting the memory reading stage (red curve).[1] Additional MLs are then performed to evaluate system stability and verify (in case there is a memory) that the memory is erased and the baseline is recovered (black and purple curves). Memory is assessed by comparing the ML heating curve before and after the RLs (red and orange curves). When the memory effect is present, the curves differ significantly near $T_R$. However, as depicted by the red curve in Figure 1c, there is no clear modification in the resistance, and the R-T curve following the RL is similar to other ML heating curves. So, there is no signature of the RRM in the heating curve.

A similar protocol modified to apply to the cooling curve is shown in Figure 2d,e, where now the RLs are performed by cooling from the T-CDW phase at 240K down to a reversal temperature. The protocol begins with a ML (from 240K to 140K) followed by four RLs (from 240K to 190.3K) and concludes with additional MLs. Now, when comparing the R-T curves of the ML before the RLs and after the RLs (red curve and orange curves, accordingly, plotted in Figures 2d and 2f) a striking difference is observed – the MIT seems to be shifted to lower temperatures. However, in the following ML (after heating to 240K and cooling again) the baseline R-T is recovered (black curve in Figure 2f). I.e., the RLs resulted in a memory effect that is erased after cooling the 1T-$TaS_2$ to low enough temperatures. This is one of the main features of the RRM effect, which was not previously reported for RLs on the cooling curve. The transition in 1T-$TaS_2$ from the N-phase to the C-phase can also be observed as a decrease in the magnetic susceptibility as function of temperature, showing a hysteresis similar to that observed in the transport measurements. We attempted to measure the RRM effect in the cooling curve of the susceptibility as well. Somewhat surprising, we could not observe a memory effect in the susceptibility measurements, see SM S5 for further details.[54] Measurements of the 1T-$TaS_2$ crystal were performed for different temperature ramp rates and for additional 1T-$TaS_2$ crystals from different batches (prepared at different times), all showing similar results, see SM section S4 for representative measurements.[54]

We emphasize that R-T behavior of the RLs is qualitatively different than that observed when measuring minor loop effects in systems with hysteresis.[42] In such systems, if the extremum temperatures are not outside the hysteretic region one would expect the phase of the system in consecutive minor loops to be driven toward its final state, i.e., toward *higher* resistance for the protocol presented here.[43,44] Whereas here we observe a *decrease* in resistance in each consecutive reversal loop, which are in essence minor loops,[1] resulting in the system's resistance perturbing outside the major loops.

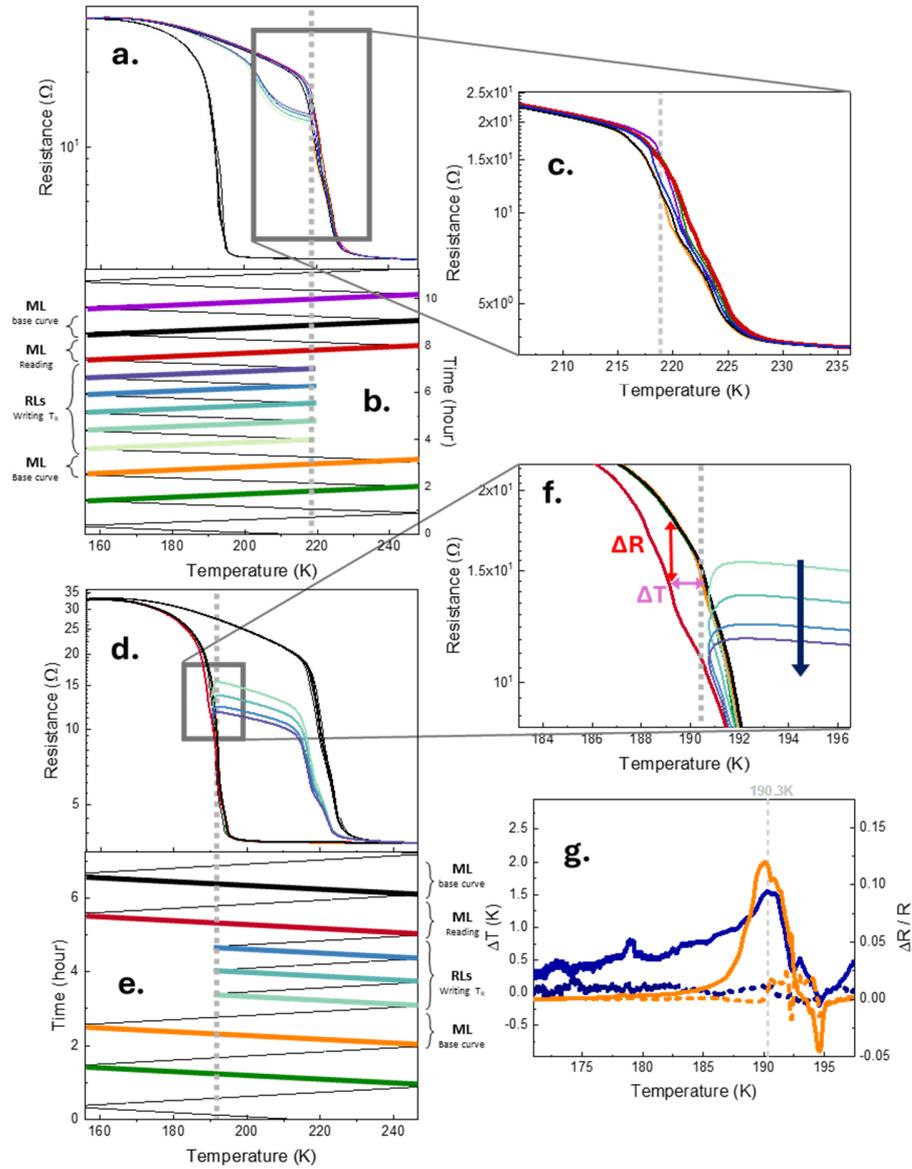

Figure 2. RRM effect measurements on a bulk 1T-TaS$_2$, for the heating and cooling branches. Heating branch: (a) R-T and (b) time vs temperature measurements of a RRM writing and reading sequence, aligned on the same temperature x axis. Each heating curve is colored using the same color coding in (a) and (b). The dashed gray line in a-c panels marks the T$_R$ of 219K. (c) A zoom-in section of (a), focusing on the T$_R$ region. The cooling curve right after the RLs is bolded. Cooling branch: (d) R-T and (e) time vs temperature measurements of RRM writing and reading sequence, aligned on the same temperature x axis. The measurements in (d) and (e) are color coded. (f) A zoom-in section of (d) around the T$_R$ region of 190.3K. ΔT and ΔR are marked. (g) Continuous lines: Plot of ΔT vs T and ΔR/R vs T for the cooling curve right after the RLs (blue and orange curves, accordingly), exhibiting a maximum at T$_R$. Dashed lines: ΔT vs T and ΔR/R for the next cooling curve (same color coding) showing the memory is erased.

Following, we demonstrate the memory properties observed in 1T-TaS$_2$ and compare them to the RRM properties reported previously in thin film TMOs.[1,4,5] Two different procedures to analyze the RRM [1,4] were previously presented: (1) calculate the normalized resistance difference (ΔR/R) vs. temperature of the ML following the RLs, and (2) calculate the temperature shift (ΔT) required to return to the original resistance at each temperature and plot it vs. temperature, see Figure 2f where both ΔT and ΔR are marked. The results of these analyses are plotted in Figure 2g, where a clear peak appears in both cases (solid orange curve for the ΔR/R analysis and solid blue curve for the ΔT analysis). The peak corresponds to the reversal temperature. Comparing analyses of the next ML (dashed orange and blue curves, following the same color coding, Figure 2g) shows the peaks have disappeared, demonstrating erasure of the memory. These features are similar to those reported previously for systems with RRM. There is, however, an interesting difference between them (in addition to the aforementioned difference that now the RLs are on the cooling curve): The magnitude of the temperature shift, ΔT, is above 1.5K, while the highest value measured previously was 0.6 K [5] and is usually around 0.15K.[4,5] We note there is an observable dip around 195K in both analyses that appears in most measurements with no relation to the reversal temperature.

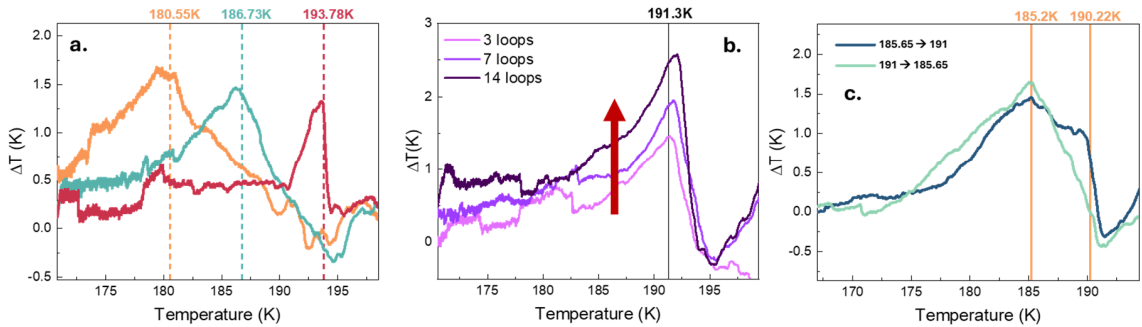

Figure 3. Different characteristics of the RRM: (a) Plot of ΔT vs T of three RRM measurements, each one of a different $T_R$, following the protocol measurement presented in Figure 2e. In all panels the perpendicular line marks the $T_R$. In **a**, the curves have the same color as their corresponding ΔT perpendicular line. All peaks agree with their corresponding reversal temperature. (b) Plot of ΔT vs T of three RRM measurements at $T_R$ = 191.58K, each measurement with different number of RLs. The red arrow indicates increase in the number of RLs. (c) Plot of ΔT vs T of protocols with two different $T_R$s, $T_R$ = 185.56 and 191K. Cyan plot for writing with decreasing $T_R$s; blue plot for increasing. Both $T_R$s are marked.

The magnitude of the RRM effect was shown to be related to the measurement of ΔT, while the ΔR/R measurement is proportional to the magnitude of the resistance change across the MIT,[5] so the ΔT analysis has a physical interpretation. Below we show various features of the RRM using the ΔT analysis, all features were observed also in ΔR/R measurements. Figure 3a demonstrates that the position of the ΔT peak follows the reversal temperature, as is the case for systems with RRM (the measurement protocol is identical to that presented in Figure 2e). Three different $T_R$s were measured, and the ΔT vs T analyses are plotted, where the reversal temperature (marked by the perpendicular lines) coincides with the corresponding peak position. Figure 3b presents the ΔT magnitude for different number of reversal loops ($T_R$=191.58K), for 3, 7 and 13 RLs. The measurement protocol is detailed in SM S5.[54] The peak position remains the same while the peak magnitude increases with the number of RLs and seems to saturate when the number of loops increases. This is similar behavior to the reported behavior in RRM systems [1,4]. In Figure 3c we show the ability to write memories simultaneously, and their dependence on the order they are written. The measurement protocol is detailed in SM S5.[54] In one case we perform 3 RLs to $T_R$ = 185.6K, followed by 3 additional RLs to a higher $T_R$ = 191K, and only then measure the memory effect (via a ML). The result is plotted as the blue curve in Figure. 3c, where 2 peaks are discerned. In the second case, the order is reversed, i.e., we first perform RLs to the higher $T_R$=191K and then the lower $T_R$= 185.6K. In this case (cyan curve in Figure 3c) the ΔT plot shows a single peak, corresponding to the effect of the second and lower $T_R$. The perpendicular lines mark the $T_R$s. This outcome indicates that (i) it is possible to write more than one memory, and (ii) reaching temperatures below the reversal temperature erases the memory set at higher $T_R$s. This is one of the principal properties of RRM systems.

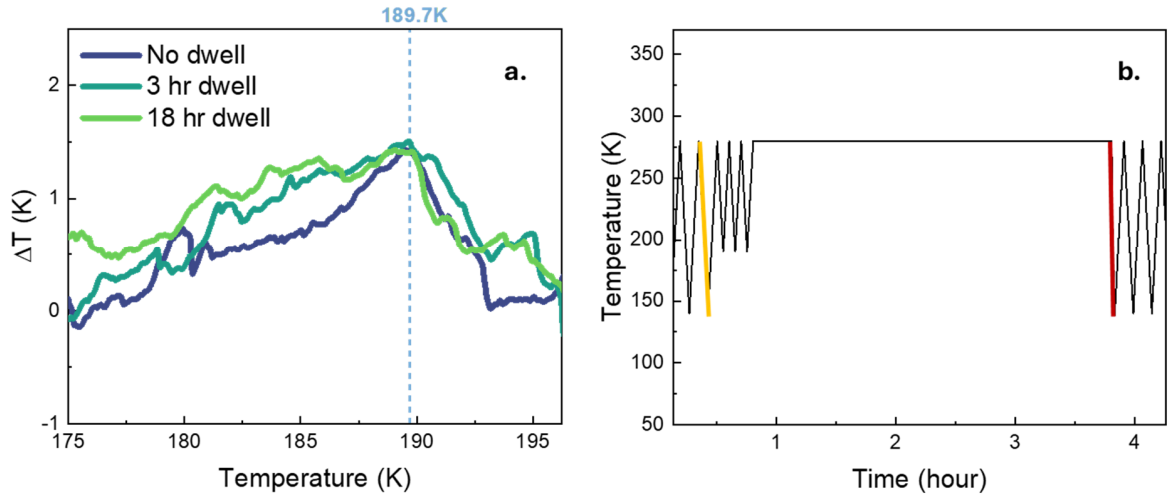

Figure 4. Three RRM measurements of $T_R$ = 189.7K following a different dwell time at 280K. (a) ΔT vs T plot of the measurements, $T_R$ is marked. (b) Temperature vs time plot of the 3-hour dwell measurement. The ΔT plot in **a** represents the comparison between the red and orange curves in **b**.

`Next, we focus on volatility of the memory effect in various scenarios. We measured the time dependence of dwelling at high temperatures after writing a memory. The system was stabilized at 280K, once for a 3-hour dwell and another time for an 18-hour dwell. The results and the measurement protocol are presented in Figures 4a and 4b, showing that the memory is not volatile on these time scales.

When we studied the RRM in oxides, cooling to very low temperatures did not erase the effect. However, in 1T-TaS$_2$ the memory is on the cooling curve, so it is reasonable that at high enough temperatures the memory will be erased. We performed the memory writing protocol using different maximum temperatures ($T_{max}$) of both the MLs and RLs, the results are presented in Figure 5. Figure 5a shows ΔT curves of RRM writing protocols with $T_R$ = 187.5K and $T_{max}$ varying between 240K to 320K. The full protocols for these measurements appear in SM section S6.[54] To follow the effect as a function of temperature we plot in Figure 5b the value of the ΔT peak as a function of $T_{max}$ (same color coding as the ΔT plots). From low temperature and up to $T_{max}$ =

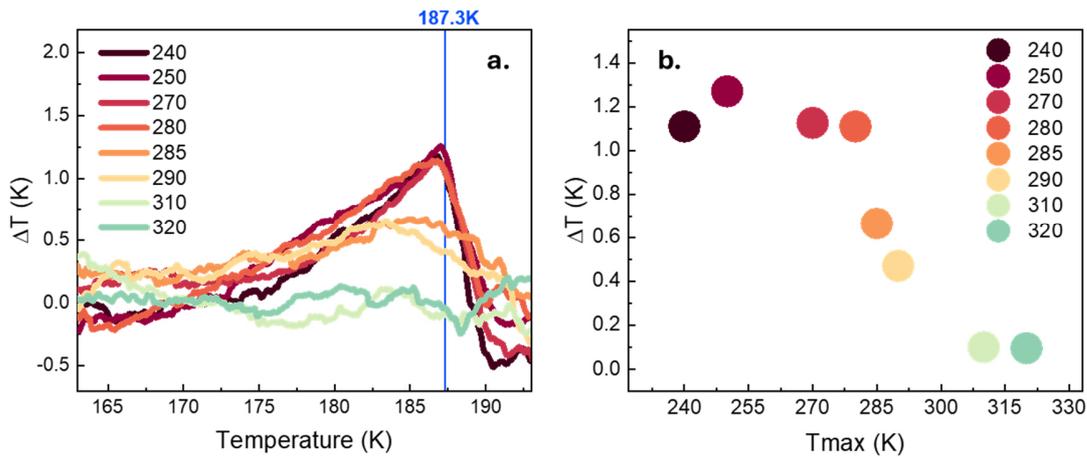

Figure 5. RRM measurements of a Tr=187.3K with different $T_{max}$. Tr is marked. (a) ΔT vs T plot of the measurements. (b) The ΔT peak as a function of $T_{max}$ of the measurement. The ΔT peaks in **b** and the ΔT plots in **a** have the same color coding.

280K, all measurements exhibit similar ΔT magnitude, around 1.2K (Note that also the peak position is the same, 187.3K). Starting at $T_{max}$ = 285K, the ΔT peak begins to decrease, and the memory disappears around ~300K. This result cannot be compared to other RRM systems and is unique (at this point) to 1T-TaS$_2$. We will return to this result in the discussion.

We investigated how dwelling at the reversal temperature affects memory properties. Figure 6 presents two measurements with $T_R$=187.5K, one without a dwell and one with a three-hour dwell at $T_R$ (detailed protocol in the SM section S8[54]). Figure 6a shows the ΔT plots, Figure 6b displays the R-T plots and Figure 6c presents resistance vs. time plots. It appears that the measurement after the dwell has a ΔT peak at a lower temperature than without the dwell. However, we can understand this result looking at Figures 6b and 6c. We observe that during the dwell the resistance of the sample increases over time (Figure 6c), indicating that the transition continues to advance, possibly due to thermal fluctuations.[45] We can estimate the effective $T_R$ from comparing the reversal loop resistance to that of the ML R-T measurement. The resistance change corresponds to a 3.2K shift in the effective $T_R$ which coincides very well with the 3.2K shift measured in the ΔT curves. Note that the resistance increase with time continues over more than a few minutes which is much longer than the temperature stabilization time in our system, so the delay cannot be attributed to some stabilization time between the sample's temperature and the temperature sensor.

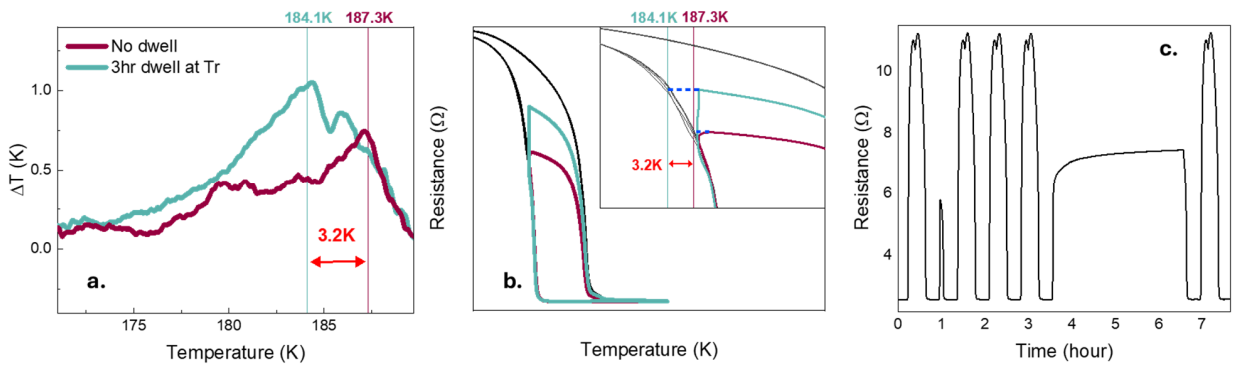

Figure 6. RRM measurement of $T_R$ = 187.3K, with and without dwell at $T_R$. (a) ΔT vs T of the measurements. Blue, with three-hours dwell at $T_R$, red, without a dwell. Solid lines mark the ΔT peaks of the measurements, with the same color coding. A shift of 3.2K between the ΔT peaks of the measurements is marked. (b) Resistance vs temperature of the measurements. In-set: A zoom-in plot of the R vs T curve, focusing on the $T_R$ region. A shift of 3.2K between the $T_R$ of the measurements is marked, corresponding to the shift in (a). (c) Resistance vs time of the measurements. It can be observed that during the whole three-hours dwell, the resistance of the sample continues to increase. The detailed protocol can be found in SM S7.

We attempted to measure the memory effect in thick flakes of exfoliated 1T-TaS$_2$. A light microscope image of a 47nm thick flake is shown in Figure 7e, see SM S9 for information on device preparation.[54] We started with a memory writing protocol in the cooling curve, as illustrated in Figure 7a,b. Here 10 RLs were conducted, with a T$_R$ of 159K and MLs with a range of 100-240K. This measurement reveals a number of interesting outcomes. First, we see that the transition in the cooling branch is steplike, while in the heating curve the transition occurs more gradually, similar to the bulk crystal. Second, the MIT transition temperature in the cooling curve is different for each cooling cycle and the hysteresis is extremely wide. Flakes with wide hysteresis were previously reported.[46,47] We could not find previous reports showing these large variations in the transition temperature between different cooling curves. The step-like transition prevents us from performing an efficient RRM writing protocol, since it requires to reverse the temperature in the mixed state, and this is next to impossible. In Figure 7b only the first RL (light purple curve) reached an intermediate mixed phase while the rest stayed metallic. Furthermore, even if we were able to reach an intermediate state, we don't have any reproducible ML to compare the curve to. We attempted to measure a number of flakes, all showing similar sharp transitions (see SM section S9[54]).

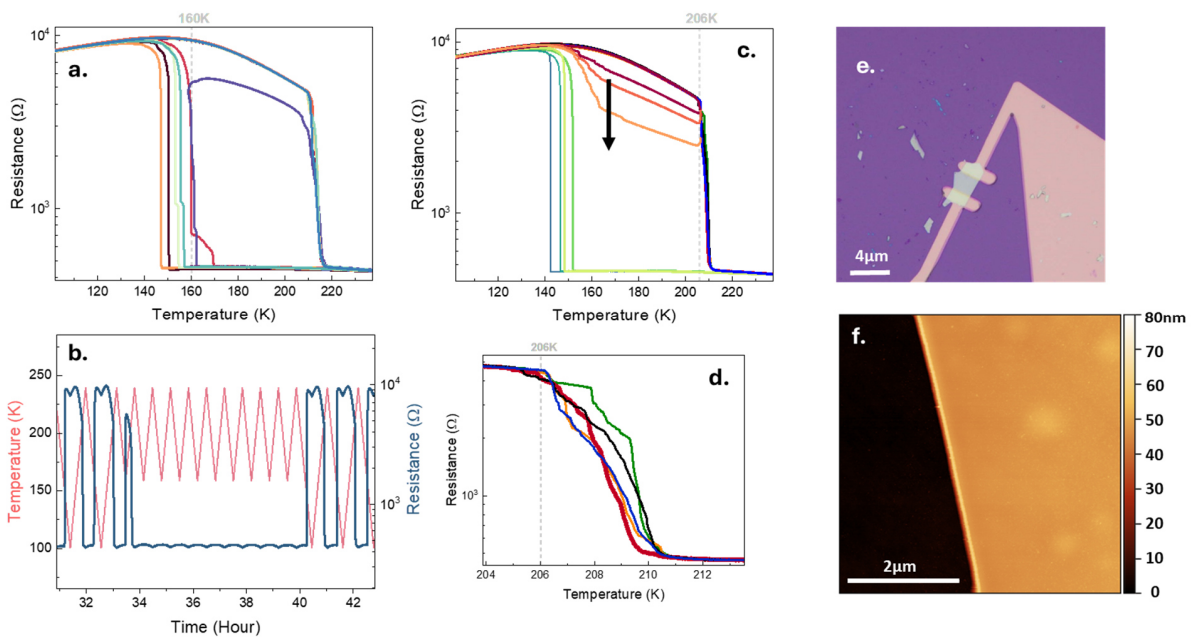

Figure 7. RRM measurement on a flake of 1T-TaS$_2$, for the cooling and heating brunches. **Cooling brunch:** (a) R vs T of a writing sequence in the cooling curve for a T$_R$ = 160K (dashed grey line). The measurement follows the measurement protocol presented in (b). (b) Temperature vs time of the measurement – red curve and left axis, and R vs time measurement – blue curve and right axis. **Heating branch:** (c) R vs T of a writing sequence in the heating curve, for a T$_R$ = 206K (dashed grey line). The measurement protocol can be found in the supplementary. The black arrow follows the decreasing trend of the reversals' resistance with consecutive RLs. (d) A zoom-in of (c), focusing on the heating curves. The red bolded curve is the heating curve right after the RLs. (e) Light microscope image of the flake and the electrodes. (f) An AFM image of the flake.

Given the better reproducibility of the heating curves we attempted an RRM writing protocol for the heating curves, shown in the R-T plot in Figure 7c. A zoom-in on the area of the MIT is plotted in Figure 7d. Each ML heating curve has fluctuation. The ML following the RLs, the red curve, represents the memory reading stage, shows an R-T similar to all other curves, indicating that no memory effect is observable. Interestingly, consecutive RL curves show a consistent decrease in resistance with each loop. This is expected for minor loops in hysteretic system, as mentioned previously [43,48], but only if the low temperature remains in the hysteretic region. If the temperature decreases below the hysteretic region, one expects the curve to repeat itself. The fact that we see the resistance decreasing between consecutive RLs indicates that there are metallic nucleation sites that were not annihilated even though the samples were cooled to 100K, well below the hysteretic region. This is an interesting phenomenon that deserves further investigation in 1T-TaS$_2$ but is beyond the scope of this letter.

## Discussion

Before we delve into the discussion of the RRM properties in macroscopic 1T-TaS$_2$, we highlight features related to the ramp-reversal experiments with the thick flakes. We observed a very sharp step-like transition in the cooling curve, and that the switching temperature varies widely with each cooling-cycle, between 144K and 169K. The heating curve, in contrast, is more smooth and does not fluctuate as much between heating cycles. This could be a result of strain from the substrate, and some stochasticity in the switching process for cooling, while there is not much strain when heating. We are not aware of this feature addressed in previous reports where flakes of 1T-TaS$_2$ were measured. As far as the RRM is concerned, we could not measure it in the flakes. This is due to the sharp MIT that changes with temperature, making it impossible to attain a phase coexistence state of the metal and insulator, which is a crucial ingredient for memory writing. I.e., the RRM cannot manifest if the single flake does not break into phase separated domains during the transition. It is plausible that measuring larger flakes (or a system with less strain where smaller flakes can break to domains) will result in the appearance of RRM also in non-macroscopic systems, from which one could learn more about the nature of the memory effect.

The results measured on macroscopic crystals of 1T-TaS$_2$ show unequivocally that a ramp reversal memory effect exists. Interestingly, some of the properties of the effect are considerably different in 1T-TaS$_2$ relative to the TMO systems where RRM was observed. In 1T-TaS$_2$ the reversal memory exists only in the cooling curve (see Figure 2), while in previous systems the memory appeared in the heating curve. The memory written at a specific T$_R$ is erased if we cool the sample to a lower temperature, in accordance with the RRM properties. Also, similar to RRM in TMOs, it is possible to write more than one memory to the system, and the magnitude of the effect increases if more than one reversal loop is performed. [1,4] Another unique property to 1T-TaS$_2$ is that the memory is retained only as long as the system is not heated above ~280K, corresponding to the transition from the T-phase to the N-phase. Meaning, if the high temperature of the major loop is in the N-phase (above the T-phase) then the memory is erased. And, of course, if the system is cooled to the fully C-phase – the memory is also erased, as shown in Figure 8a. This indicates that the memory is maintained by a feature related to the T-phase and its phase boundary with the C-phase.

A plausible scenario is that during a ramp reversal, when the system reaches T$_R$ and is in the phase-coexistence state (of T and C phases), the stacking order of the area near the phase boundaries is altered. It is known that stacking order can change the electronic properties of 1T-TaS$_2$ [37,39,49,50], and in this case, it causes the Tc of the affected areas to decrease. This new stacking order remains when the system returns to the T-phase and is erased only when all the T areas of the phase boundaries transform to a C-phase (what we refer to as healing the scars [1]), or when the system is heated to the N-phase, which also erases any property related to the C-phase.

It is known that the 1T-TaS$_2$ system can be driven out of thermal equilibrium into temporarily metastable states by pulses of light or current. [35,37,51,52] We suggest that the affected areas at the phase boundaries are also driven to some sort of local metastable state that maintains the low resistance state in a similar manner .[37] The origin must be different, of course, since in all studies the hidden state was reported to disappear above 70K to 120K [33,52,53], while our measurements are above 250K. Unlike previous reports of hidden states in 1T-TaS$_2$, where the metastable states were achieved by exciting the system out of equilibrium, here it occurs through thermal equilibrium, requiring only heating and cooling of the sample. The suggested explanation for the scar formation is general and heuristic in nature, requiring development of a theoretical microscopic explanation. Solving this issue will require further studies with different measurement techniques, such as local probe measurements of the RRM.

An additional difference between the RRM in 1T-TaS$_2$ and TMOs, is the magnitude of the ΔT peak. The ΔT peak magnitude measured in previous system was ~0.15K in VO$_2$ on c-cut and r-cut sapphire, 0.2K in NdNiO$_3$, and a maximum value of ΔT~0.6K was observed for VO$_2$ deposited atop SiO$_2$. [5] Here, we measured a ΔT peak surpassing 2.5K (for a few RLs). Note that this value indicates how much the transition temperature is shifted due to the memory effect, and a 2.5K shift is a substantial effect. We identify two parameters that can affect the magnitude of the peak. The first one is that the local temperature shift at the phase boundaries is larger in TaS$_2$ during the 'writing' stage, resulting in a shift in the measured R-T curves. It was shown [5] for VO$_2$ thin films that if they are less strained by the substrate, i.e., when the film is not lattice matched, then during the ramp reversal, the phase-boundary distortions (scars) are more free to reorganize. This resulted in more stable scars, that require larger energy to be released, i.e., a large $\Delta T$ and a larger RRM effect. Here, 1T-TaS$_2$ is a 3-dimensional macroscopic Van der Waals layered material. It is reasonable that flakes at the phase boundary of the phase-separated state could slip relative to one another, resulting in high-energy phase-boundary scars. [5]

A second source is related to how the local properties of 1T-TaS$_2$ change spatially. According to the RRM heuristic model the memory develops in the phase boundaries of the phase-separated domains (here between the T- and C-phases). If the transition temperatures of nearby domains are correlated, i.e. the spatial change of T$_C$ is slow, then the ΔT peak will be smaller than if the neighboring domains are not correlated and may have very different switching temperatures. The reasoning is that if the domains are less correlated then more T-phase islands will appear, resulting in more phase-boundaries during the reversal, and then more areas where the transition temperature changes, leading to a larger effect.

To demonstrate this, we performed numerical simulations where we include a "correlation length" parameter, $\xi$. The model simulates the nucleation and growth of the C-phase in the T-phase as the system is cooled. As the correlation length increases, the Tc of more distant sites will be correlated. The local $T_C$ and the correlation length are controlled by randomly assigning $T_C$s from a gaussian distribution to a specific percentage of sites, and then smoothly changing the $T_C$ between these sites. The correlation length is then defined as the average distance between uncorrelated sites, so that $1/\xi = 1$ indicates adjacent sites are uncorrelated and as $\frac{1}{\xi} \to 0$ means the correlation length $\xi \to \infty$ and all sites are correlated. See Figure 8b for an example of a $T_C$ map for a realization with $\xi = 3.16$ sits. In each realization of the simulation the system follows a standard RRM protocol and calculates the ΔT plot between the "reading" curve and "base" curve. For more details on the numerical simulation see SM S10.[54]

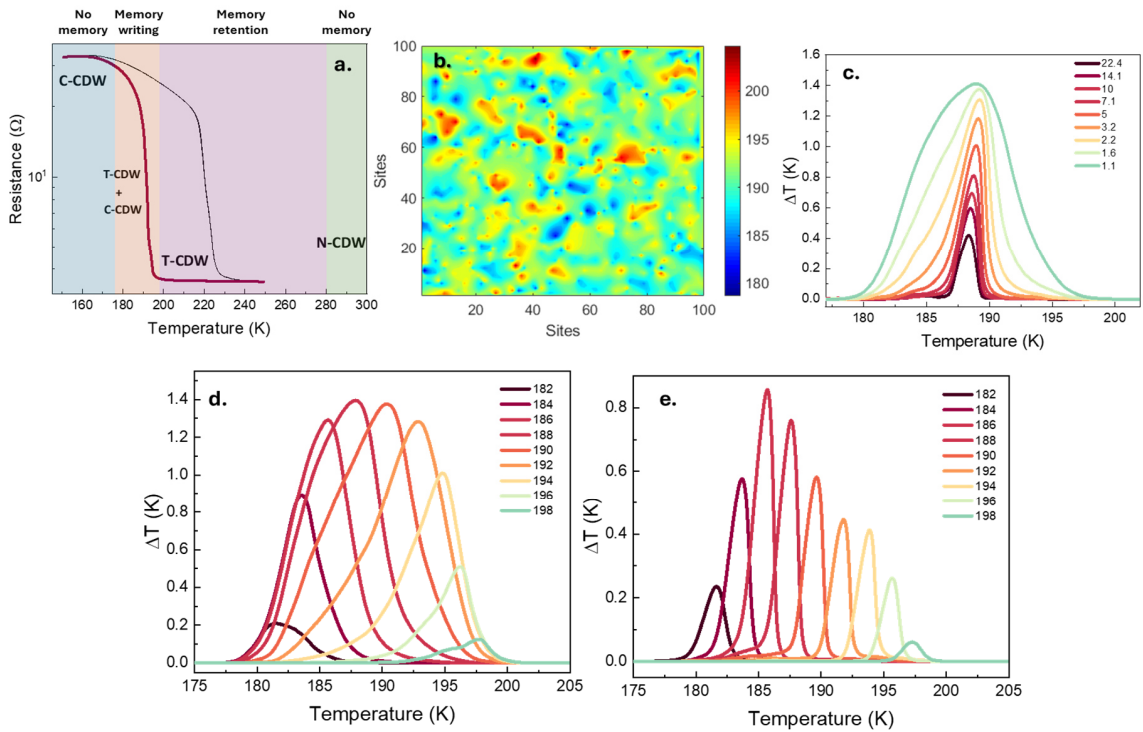

Figure 8. Numerical simulation of the RRM effect in 1T-TaS$_2$. (a) Schematic of the RRM memory writing stages, different CDW phases, and resistance as function of temperature for a bulk TaS$_2$. (b) T$_c$ distribution map for $\xi = 3.16$. (c) ΔT vs temperature calculation for T$_R$=189K plotted as a function of different $\xi$ values, appearing color coded in the legend. (d) ΔT vs temperature simulation for different T$_R$ with $\xi = 1.3$. (e) ΔT vs temperature simulation for different T$_R$ with $\xi = 10$. The local Tc shift in d. and e. is 1.5K.

Results for simulations with different $\xi s$ are plotted in Figure 8c. The $\Delta T$ vs temperature curves of an RRM protocol for $T_R=189K$ are shown, the correlation length is marked in the legend, ranging from $\xi = 22.4$ sites, to short correlation length $\xi = 1.1$, see legend. The local $T_C$ shift for the sites at the phase-boundaries during the reversal were set to 1.5K. As the correlation length increases the peak increases, starting from 0.4 and growing toward the local Tc shift of 1.5. In addition, the width of the $\Delta T$ also grows. This demonstrates our claim that the measured $\Delta T$ will be larger as the correlation length becomes smaller. We attain additional insight from the simulations in Figure 8d and 8e, where we simulate the $\Delta T$ vs temperature for different reversal temperatures (see legends) and two different correlation length, Figure 8d for $\xi = 1.3$ sites and 8e for $\xi = 10$ site (the local Tc shift is 1.5K). In both cases the $\Delta T$ peak value is larger at the center of the transition, where many sites are switching so there are more scars, and smaller toward the beginning and end of the transition where there are fewer scars. However, there are notable differences. First, the magnitude and the width of the $\Delta T$ peak is larger for the shorter $\xi$ for all reversal temperatures, as explained above and shown in Figure 8c for a single $T_R$. Second, for short $\xi$ there is a wide range of reversal temperature, between 185K and 195K, that have a similar $\Delta T$ magnitude, while for longer correlation length the peak magnitude changes with each reversal temperature.

Considering the experimental results, which show that the $\Delta T$ is larger and wider in TaS$_2$ as compared to TMOs, and that the magnitude is similar for different reversal temperatures (compare Figure 3a with reference [5]), we hypothesize that the correlation length in TaS$_2$ is shorter than in TMOs. This is plausible since the TMO were thin granular films, where each grain can be a nucleation site, and the composition of adjacent grain should, in general, have similar characteristics of oxygen vacancies, strains, etc., resulting in a similar transition temperature that changes only in grains further away. Note that the correlation length should be normalized to the average size of the domains that are switching of course. The TaS$_2$, is a single crystal, so the domains with different transition temperatures are probably larger and could have much less correlations between them.

# Conclusion

In summary, we have measured the presence of the RRM in bulk 1T-TaS$_2$, demonstrating the generality of the ramp reversal effect. While 1T-TaS$_2$ has the main RRM properties as previously demonstrated in TMO systems, a notable difference is that the effect exists in the cooling branch, and not in the heating branch. We show the mechanism of the RRM is related to the phase coexistence between T- and C- CDW phases, explaining why it can only be observed on the cooling branch. However, a microscopic mechanism for memory writing in this system is still lacking. We hope this study will encourage further experimental and theoretical research into this intriguing phenomenon. Importantly, the discovery of RRM in an additional system so different from TMOs emphasizes the universality of this memory effect. Our results suggest that RRM could be a more general phenomenon in correlated materials with a phase transition occurring through spatial phase separation. Meaning, it should exist in systems with ferromagnetic, ferroelectric and other types of phase transitions, where the ramp-reversal of a driving force can lead to emergent memory phenomena in these correlated systems.

# Methods

**Resistance vs Temperature measurements.** Silver paste was used to connect wires to the 1T-TaS$_2$ bulk sample. The sample was cooled by our home-made cryostat with a temperature range of 77K to 370K. The resistance was measured by a Keithley 2400, with current source of 4mA. The temperature was controlled by a Lakeshore 335 and was ramped with 5 kelvin/min. Both 2 and 4 probe measurements were performed. We compared the two and prove that the 2 probe measurements are reliable, see SM S3 for more information.[54]

**Flakes fabrication and measurements.** 1T-TaS$_2$ flakes were fabricated by using an exfoliating technique. 1T-TaS$_2$ bulk was pressed to a tape, and then the tape was pressed onto a SiO$_2$ 300nm/Si substrate, leaving flakes on the surface. Then, Au contacts were added by a photolithography mask planned for each flake. The flakes were measured by the same setup measurement of the bulk, with a voltage source of 1mV.

**Numerical simulations.** The numerical simulations were performed using Matlab software. A resistor network with local resistances changing with temperature was defined, and a set of Kirchoff equations were written. The total resistance of the network was found by inverting the matrix and solving the total current for a specific voltage drop used as the boundary condition. A full explanation of how the resistors local resistance change with temperature are defined appears in SM S10.[54]

## ASSOCIATED CONTENT

Supporting Information including additional measurements, detailed measurement protocols, and detailed explanation of the numerical simulation is available. This material is available free of charge via the Internet at http://pubs.acs.org.

## ACKNOWLEDGMENT

AS acknowledges the support of the Israel Science Foundation No. 1499/23 and MOST No. 6068 for their support of this work.

## ABBREVIATIONS

RRM – Ramp reversal memory

MIT – Metal insulator transition

TMO – Transition metal oxide

SM – Supplementary material

ML – Major loop

RL – Reversal loop

$T_R$ – reversal temperature

CDW – charge density wave

# Supplementary Information for Ramp reversal memory in bulk crystals of 1T-TaS$_2$

Avital Fried[1,2], Ouriel Gotesdyner [1], Irena Feldman[3], Amit Kanigel[3] and Amos Sharoni[1,2,*]

[1] Department of Physics, Bar-Ilan University, 5290002 Ramat-Gan, Israel

[2] Bar-Ilan Institute of Nanotechnology and Advanced Materials (BINA), 5290002 Ramat-Gan, Israel

[3] Department of Physics, Technion-Israel Institute of Technology, Haifa, 32000, Israel

*Amos.Sharoni@biu.ac.il

## Contents

# 1. The RRM effect – writing example in VO$_2$ thin films

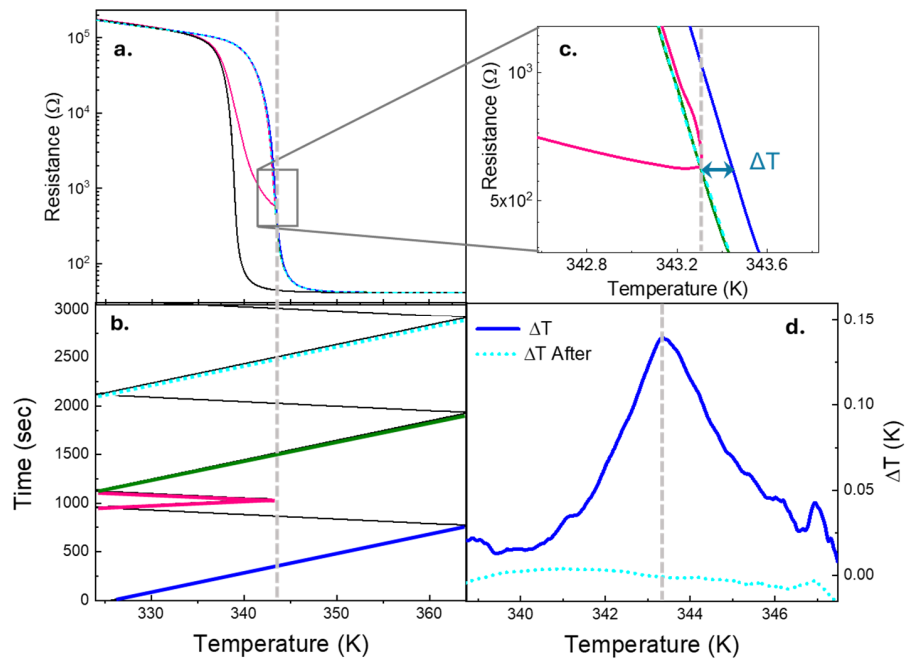

**Figure S1. RRM in VO$_2$ thin films.** a. R-T of the MLs and RL, color coded to match the heating and cooling protocol plotted in b. T$_R$ is marked by dashed grey line. c. Zoom in on the region of the reversal temperature, showing the change in resistance following the RL and recovery of the original R-T in the following measurement. ΔT is marked in the image. d. Plot of ΔT vs temperature right after the RL (blue peak) and in the following heating measurement (cyan dashed line, no peak).

Figure S1 presents an overview of the ramp reversal memory (RRM) writing protocol, as measured in a thin film of a 180nm VO$_2$ sample deposited on a c-cut sapphire. VO$_2$ is a transition metal oxide with a metal insulator transition at 342K. It has all the ingredients deemed necessary to show RRM, as recapped in the main paper. The memory writing protocol is exemplified in the temperature vs time plot shown in Fig. S1b, and the corresponding resistance vs temperature (R-T) curves shown in Fig S1a. The color coding of the curves is the same in both graphs. The protocol begins with the sample at the insulating state, at 328K. Then, the sample is heated to 358K to the fully metallic state, and then cooled back to the insulating state (green and black curves in S1a and S1b), creating a full hysteresis loop referred to as a major loop (ML). Then, the sample is heated to an intermediate temperature in the phase-coexistence state (where both metallic and insulating domains coexist and are spatially separated). The intermediate temperature is referred to as the "reversal temperature", T$_R$. Following, the film is cooled back to 328K (pink curves in 1a-c). This is the memory writing stage which we refer to as the reversal loop, RL. Fig. S1c presents a zoom-in view of Fig. S1a, focusing on the $T_R$ area. Following is the "reading" stage, where the sample is heated to 358K (blue curve) and back to 328K (black curve). When comparing the heating curve right after the RL (the blue curve) with the first heating curve (green curve), one can clearly observe the curves are significantly different around $T_R$. We analyze the data by plotting the difference between the curves by calculating for each temperature what is the delay in the temperature needed to return to the original R-T curve, referred to as $\Delta T$, see Fig. S1c. In essence we measure the temperature difference between two points with the same resistance on each of the curves. When plotting $\Delta T$ in Fig. S1d, a peak appears corresponding to the reversal temperature. This is the memory. In the consecutive heating curve this peak disappears, plotted as the dashed line.

## 2. Crystal fabrication

The crystals were prepared by the chemical vapor transport (CVT) method, as described in [1]. This process resulted in crystallites a few millimeters in length, but thinner along the c-axis of the van der Valls materials. An XRD measurement of the main batch used in this research is plotted in Fig. S2. The single crystal nature of the sample is evidenced by the sharp peaks of only the 00n direction, which is the direction measured.

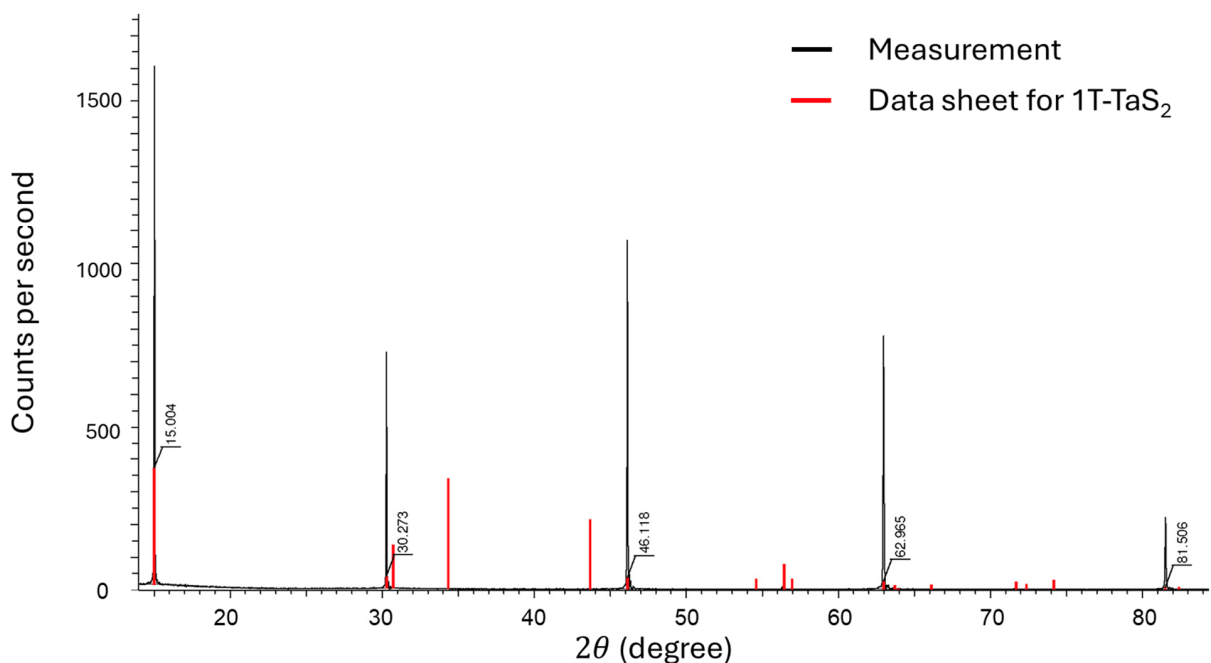

**Figure S2. XRD of 1T-TaS₂.** $\theta - 2\theta$ XRD measurement of sample in the c-direction. Only 00n peaks are evident. The MSDS peaks appear as red lines.

## 3. Transport measurements and comparing two- and four-probe measurements

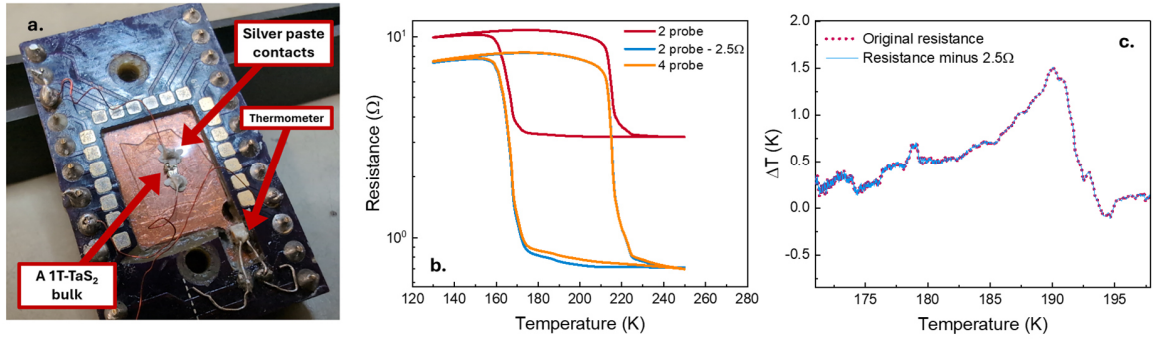

**Figure S3 Transport measurements.** (a) An image of the 1T-TaS$_2$ bulk crystal connected with silver paste to a sample holder. The crystal was glued to a thin glass slide using g-varnish, providing insulation from the sample holder. The glass is attached to the sample holder with thermal grease. The Pt temperature sensor can be seen.
(b) R-T measurements of a1T-TaS$_2$ bulk by 2-probe (orange) and 4-probe (red) methods. The blue curve is the plot of the 4-probe measurement after subtraction of 2.5Ω. (c) ΔT vs temperature plot of the RRM protocol shown in Figure 2, for the original cooling resistance curves (dashed pink) and for the cooling resistance curves after a subtraction of 2.5Ω.

For the resistance measurements of bulk TaS$_2$, silver paste was used to connect the wires to the sample. An image of a device is provided in Fig. S3a. We performed both 4-probe and 2-probe measurements and compared their results. Figure S3b shows the R-T measurements for the same sample: the red curve represents the 2-probe measurement, and the orange curve represents the 4-probe measurement. When the wire's resistance (2.5 Ohm) was subtracted from the 4-probe measurement the two curves overlap. Fig. S3c compares the ΔT analysis of the 2-probe measurement with and without subtracting the wire resistance. Since ΔT measures only the shift in the transition temperature, these plots should be identical, and Fig. S3c shows that indeed they are. Due to the difficulty in connecting 4 wires to most samples, since they are small, we used 2-probe setup in most of the measurements.

## 4. RRM measurements at different ramp rates and from a different 1T-TaS$_2$ batch

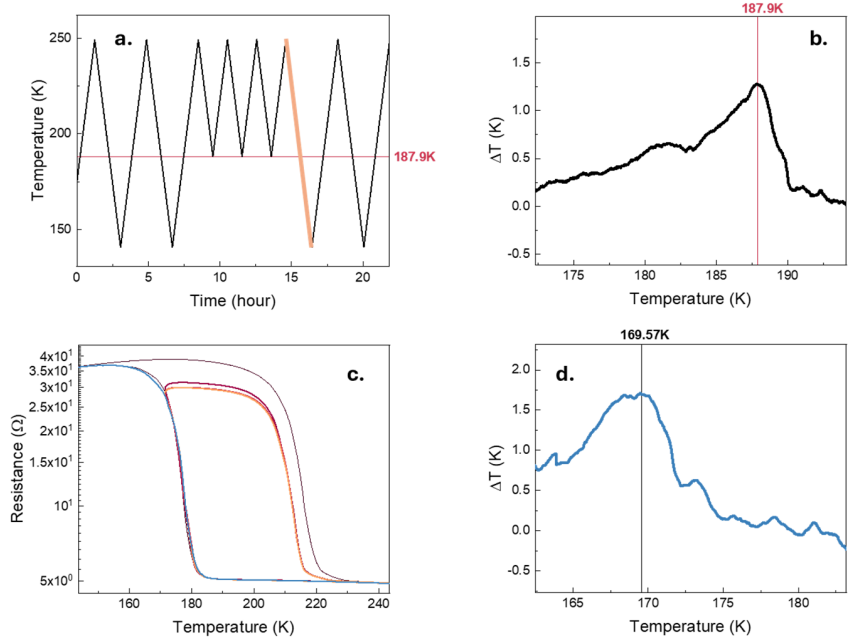

**Figure S4.** (a) Slow speed temperature vs time of a RRM protocol with three RLs with $T_R$=187.9K ($T_R$ is marked). The ramp rate of this measurement is 1 K/min. The reading stage of the memory is marked in orange. (b) $\Delta T$ vs temperature calculation of the orange curve in a. $T_R$ is marked. (c) Resistance vs temperature measurement of an additional 1T-TaS$_2$ bulk of another batch. The measurement follows a regular RRM protocol of three RLs with $T_R$=169.57K. $T_R$ was calculated the same way as in Figure 6. The cooling curve right after the RLs is colored in blue. (d) $\Delta T$ vs temperature calculation of the "reading stage" in c (blue cooling curve).

To investigate whether the measurement rate influences the RRM characteristics, we measured the RRM effect in bulk TaS$_2$ using a slower ramp rate of 1 kelvin/min. Figure S4a shows the temperature vs time plot for this slower measurement, including three RLs and $T_R$=187.9. The cooling curve right after the RLs is marked orange. Fig. S4b presents the $\Delta T$ vs T plot of the RRM, revealing a peak at $T_R$ with a value similar to those observed at higher ramp rates, compare to the measurement shown in Fig. 2 of the main article, where a rate of 5 kelvin/min was used. Thus, for these rates there is no difference. We used the faster rate for the measurements.

We measured the RRM effect in a crystal from a different batch of 1T-TaS$_2$. A measurement from this batch is shown in Fig. S4c,d. Fig. S4c shows the R-T measurement for a standard RRM measurement protocol with three RLs and $T_R$=169.57K and Fig. S4d is the $\Delta T$ vs T plot. The $\Delta T$ exhibits a peak at $T_R$, with a value similar to those observed in the first batch (compare to Figure 2 and 3 in the main article).

## 5. Susceptibility measurements of TaS$_2$: RRM protocol.

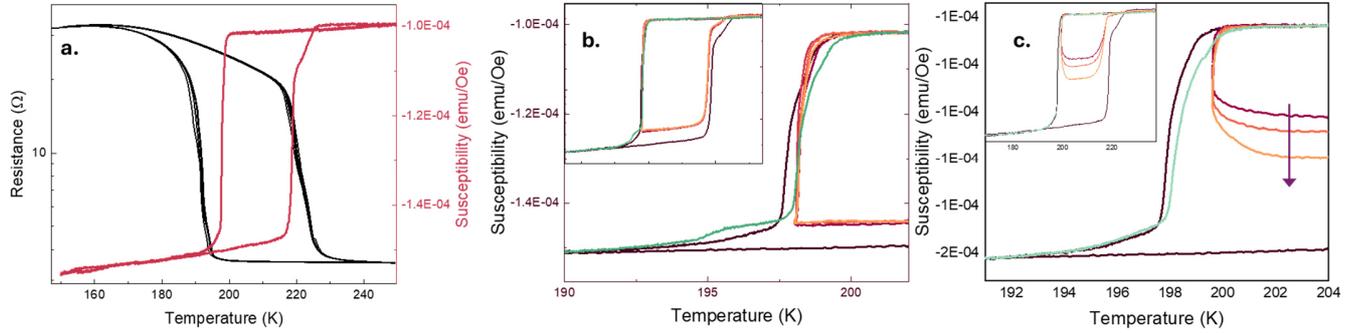

**Figure S5. Susceptibility measurements of 1T-TaS$_2$**. (a) Resistance (black) and susceptibility (red) measurements of the same 1T-TaS$_2$ bulk. (b) A susceptibility measurement of the crystal, following a standard RRM measurement protocol with three RLs and T$_R$=198K. The cooling curve right after the RLs is marked in green. Inset shows the full temperature range of the susceptibility measurement. (c) Same as b but with T$_R$=199.6K. The cooling curve right after the RLs is marked in green.

TaS$_2$ also exhibits a transition in magnetic susceptibility when it goes through the metal insulator transition at 195K, showing a hysteresis.[2,3] We attempted to observe the RRM effect in susceptibility measurements, and preformed susceptibility measurements on a 1T-TaS$_2$ bulk sample following a general RRM protocol. Fig. S5a depicts the susceptibility measurement vs temperature, showing a hysteresis with relatively sharp transitions in cooling and heating, red curve, together with the R-T measurement of a sample from the same batch. Measurements were performed separately. Notably, the transition in resistance and susceptibility during cooling aren't synchronized. The susceptibility transition is sharper and occurs at somewhat higher temperatures, while the resistance transition begins only after the susceptibility transition is complete. Additionally, the hysteresis width is different. So, while the difference in transition temperature could be due to differences in the temperature sensing apparatus, this cannot accommodate for the difference in the width of the hysteresis. We are not aware of this observation been reported in previous papers. This could be an interesting point to explore.

The RRM protocol for the susceptibility is plotted in Fig. S5b and S5c for two reversal temperatures, T$_R$ = 198K and 196K, correspondingly. In each case three RLs were performed. The main figure focuses on the cooling branch, and in each figure the inset is of the full loop. For T$_R$ = 198 the cooling curve right after the RLs (green curve) is lower than the cooling curve before the RLs (dark red curve) across most of the transition, but near 196K, the curve shows higher values, similar to the increase observed in the RRM effect near $T_R$. However, this effect was not consistent, as can be seen for the measurement of $T_R$ = 196, shown in S5c. First, the reversals curves exhibit a decreasing trend with each subsequent reversal loop, showing lower susceptibility values for the same temperature. This is indicative of a nucleation and grow behavior, where the nucleation sites have not been annihilated,[4] and is opposite to the trend we see for the RRM effect. Second, the cooling curve right after the reversal loops (green curve in Fig. S5c) is consistently lower than the unperturbed cooling curve, dark red curves in Fig. S5c. Thus, interestingly, we cannot observe the RRM effect in the susceptibility measurement. This may indicate a separation between the magnetic transition and electronic transition when reversal loops are performed, but additional research is needed to verify this statement, which is beyond the scope of this study.

## 6. RRM protocols used for Figure 3 in main paper

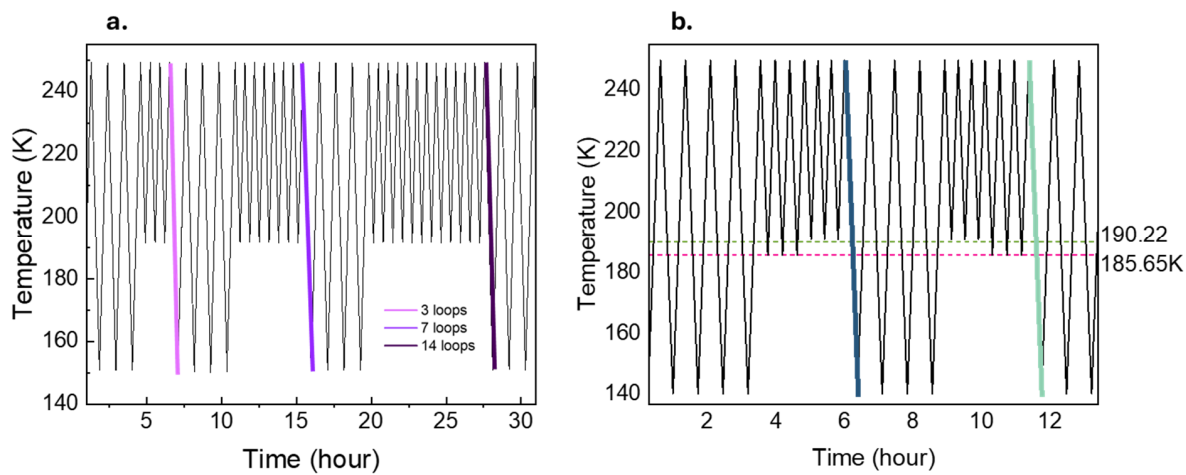

**Figure S6.** RRM protocols of Figure 3. (a) Temperature vs time for the three measurements of different number of loops shown in Figure 3b. The "reading stage" of each measurement is mark in different color, with the same color coding as in Figure 3. (b) Temperature vs time for the two measurements of two different $T_{RS}$ shown in Figure 3c, each one with different $T_{RS}$ order. The "reading stage" of each measurement is marked with the same colors as in Figure 3. The $T_{RS}$ are marked in dashed lines.

Time vs temperature measurements for the ΔT plots in Figure 3b,c main paper are presented. The cooling curves have the same color coding as the ΔT plots in Figure 3 of the main paper. In Fig. S6b, the two $T_{RS}$ are marked by the dashed lines.

## 7. RRM protocol of measurements for Figure 5 main paper

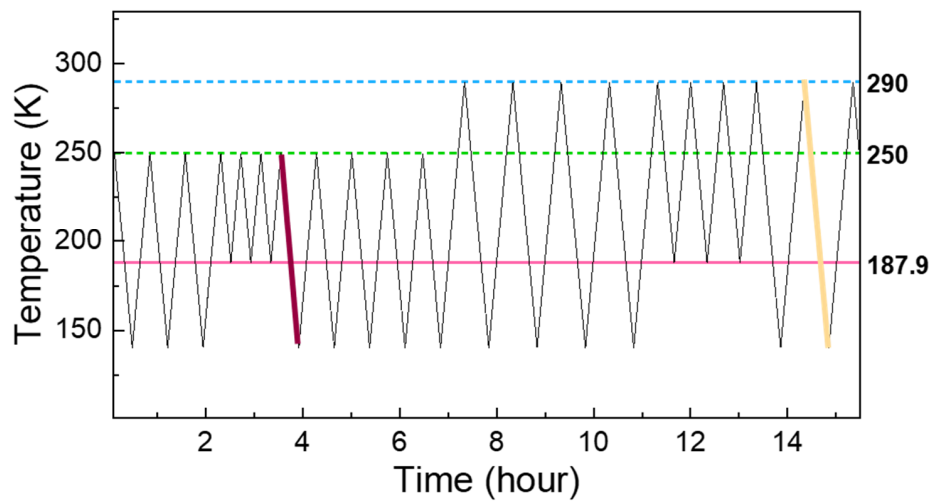

**Figure S7.** Temperature vs time for two of the different $T_{max}$ measurements shown in Figure 5 main paper: $T_{max}$ = 250 and 290K. $T_R$ of the two measurements is the same, $T_R$ = 187.9K and marked as pink line. The "reading stage" of each measurement is mark in different colors.

The Time vs temperature protocol of the measurements shown in Figure 5 are plotted. The $T_{max}$ shown are for 250K, 290K.

## 8. RRM protocol for Figure 6 main paper

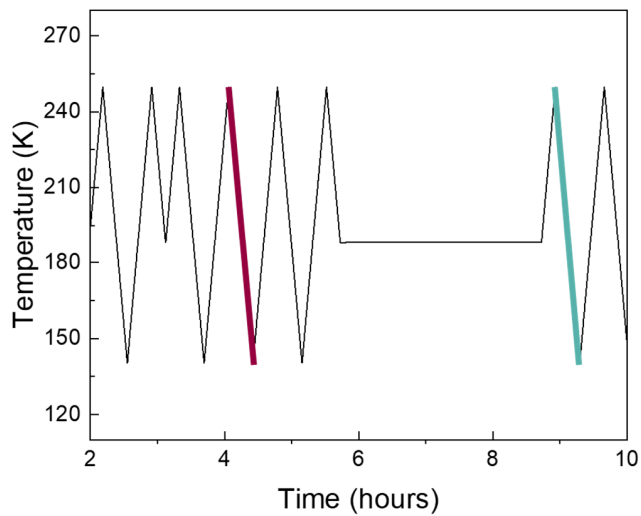

**Figure S8.** Temperature vs time for the two measurements shown in Figure 6. The "reading stage" of each measurement is marked with the same colors as the colors of the matching ΔT measurement in Figure 6a.

Time vs temperature plots of the measurements shown in Figure 6 main paper, both with the same $T_R$, one without a dwell and the other with a dwell of 3 hours at $T_R$. The cooling curves have the same color coding as the matching ΔT plots in Figure 6.

## 9. Additional flake measurements

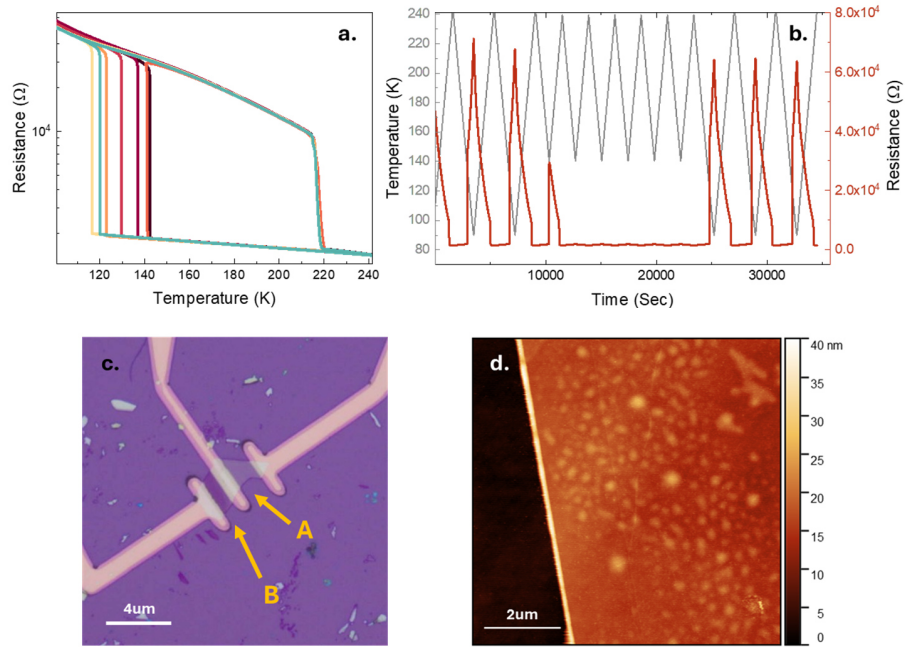

**Figure S9.** Additional flake measurements. (a) Resistance vs temperature of a 18nm flake. The measurement follows the RRM protocol shown in b, with five reversals and $T_R$ = 140.5K. (b) Temperature and resistance as a function of time of the measurement in a. It is seen that a transition occurred only through the first reversal, for the rest the system stayed in the metallic state. (c) A microscopic image of the flake and contacts. The contacts are arranged that two areas of the flakes can be measured separately. The measurements shown here are of area A. (d) An AFM image of the flake.

The $TaS_2$ flakes were prepared using the exfoliating technique. Thin layers were separated from the bulk material using tape, then transferred onto a 300nm $SiO_2$/ Si substrate by pressing the tape onto the substrate. Contacts were added through photolithography, with 15 nm of Cr and 70 nm of Au.

Measurements of an additional flake with a thickness of 18nm show results consistent with those presented at Figure 7 of the main paper. Figure S9a shows the R-T measurement of the flake, following the RRM protocol measurement shown in Fig. S9b. In S9b, both temperature vs time (grey) and resistance vs time (red) are plotted. Here $T_R$=140.5K and 6 RLs were performed. The transition in the cooling branch is steplike, the transition temperature varies with each ML cycle, and the hysteresis is notably wide (80K to 100K). Figure S9c shows a microscope image of the flake. The measurements shown here are from device A on the flake. An AFM of the flake edge from which the thickness was measured is shown in Fig. S9d.

# 10. Numerical simulation – details

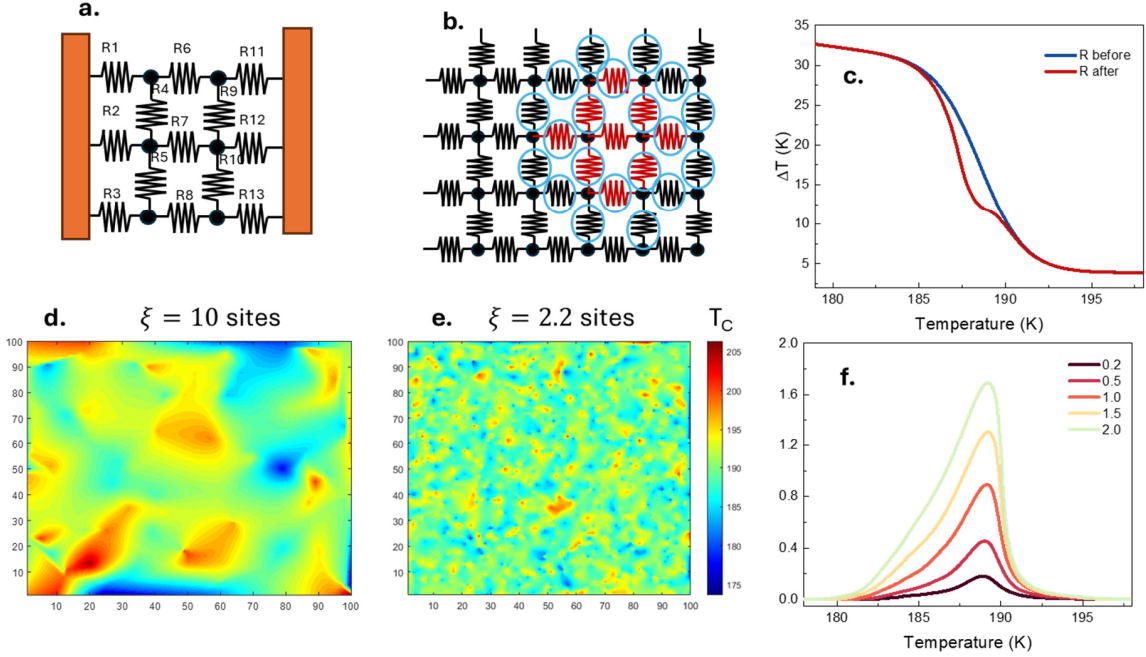

**Figure S10. Numerical resistor networks**. (a). a depiction of the resistor network with 4 resistors connected to each node. The edges are conducting. (b). Map of switched (red) and unswitched(black) resistors at the reversal temperature. Edge resistors are marked by blue circles. (c). and (d). Maps of local Tcs for 2 realizations, one with a long coherence length of 10 nodes, and the other for a short $\xi$ of 2.2 nodes. (e) A resistance vs temperature curves for the RRM protocol, showing the resistance curves before and after the reversal. Here, $\xi = 10$ nodes ($1/\xi = 0.1$).

The computational model consists of a 2D resistor lattice network with $n \times n$ nodes ($100 \times 100$ nodes were used in all simulations). Thus, the number of resistors (links) is $2n^2$. See Fig. 10a for a depiction of a smaller network (3x2). The resistor network is connected at its edges to a low resistance conductor (orange in the image). Each resistor can be in one of two states, a higher insulating resistance R$_{HIGH}$, and a lower metallic resistance R$_{LOW}$. For the high resistance we used a temperature dependence $R_{HIGH} = R_I \cdot \exp\left(-\frac{T}{\gamma}\right)$ to imitate the temperature behavior of the high resistance state. The constants were chosen so the R-T will follow the experimental data.

Each resistor has its own critical temperature for the transition between the low and the high resistance state. We want the resistors to change continuously over space with some controlled correlation length $\xi$, thus forming nucleating islands that grow. To achieve this, we used the following procedure. First, we choose the number of nodes, N$_{Tc}$, that will receive a random transition temperature from a gaussian distribution with $T_{C,avg}$ and $\sigma_{T_C}$ distribution width. These values were also chosen so the numerical R-T and the experimental one will be similar. The correlation length is defined now as $\xi = \sqrt{\frac{n^2}{N_{Tc}}}$. Next, we build a spatially 2D linear function of Tcs from the 2D distribution of Tcs and calculate for each node its Tc. Examples of Tc maps for a long and short correlation length appear in Figures S10c and S10d, accordingly. Finaly, the Tc of each resistor was defined as the average critical temperature of its two nodes.

Now we are ready to perform a RRM protocol. For a given realization we first calculate the R-T by discretization of the temperature steps (usually ~0.2 degrees) and at each temperature calculating the resistance of the network by standard matrix inversion, see Fig S10c where a numerical R-T is shown. Next, the effect of the reversal loop was performed as follows. At the reversal temperature, $T_R$, we have a map of switched resistors and unswitched resistors. See Fig. S10b, where the black resistors are unswitched and the red ones are switched. Now all the resistors that are on the edge have a chance of their transition temperature being increased. An edge resistor is defined as a resistor that another resistor in an opposite state is connected to one of its nodes. All the edge resistors are marked by blue circles in Fig. S10b. We chose that all resistor that have a neighboring different resistor – their Tc will change, but other conditions could be chosen. If the Tc of the resistor is changed then it is reduced by a constant value of $\Delta T_{loc}$.

After this stage is completed, we perform another R-T measurement of the network with the new local transition temperatures. The calculation of $\Delta T$ vs temperature is done using the two R-T curves. For each reversal temperature we average over 20-30 realizations.

The values used in the simulations appear in the table:

| Variable | $n$ | $R_{LOW}$ | $R_I$ | $\gamma$ | $T_{C,avg}$ | $\sigma_{T_C}$ | $\xi$ | $\Delta T_{loc}$ |
|---|---|---|---|---|---|---|---|---|
| Value | 100 | 3.6 | 200 | 100 | 190 | 4 | 14.1-1.05 | 1.5 |

The range of temperatures simulated was 240 to 160.